# Testing GitHub projects on custom resources using unprivileged Kubernetes runners

Igor Sfiligoi

University of California San Diego, USA, isfiligoi@sdsc.edu

Daniel McDonald

University of California San Diego, USA, danielmcdonald@ucsd.edu

Rob Knight

University of California San Diego, USA, robknight@ucsd.edu

Frank Würthwein

University of California San Diego, USA, fkw@ucsd.edu

GitHub is a popular repository for hosting software projects, both due to ease of use and the seamless integration with its testing environment. Native GitHub Actions make it easy for software developers to validate new commits and have confidence that new code does not introduce major bugs. The freely available test environments are limited to only a few popular setups but can be extended with custom Action Runners. Our team had access to a Kubernetes cluster with GPU accelerators, so we explored the feasibility of automatically deploying GPU-providing runners there. All available Kubernetes-based setups, however, require cluster-admin level privileges. To address this problem, we developed a simple custom setup that operates in a completely unprivileged manner. In this paper we provide a summary description of the setup and our experience using it in the context of two Knight lab projects on the Prototype National Research Platform system.

## 1 INTRODUCTION

The Knight lab [1] maintains several bioinformatics projects using **git** as the version control system. While git does not require a central repository, an authoritative location is needed for any serious endeavor, so GitHub [2] has been chosen due to its popularity, ease of integration with external users, and great pull request management. The integrated testing environment capabilities through **GitHub Actions** [3] is a further benefit. GitHub is free to use, which is a valuable benefit, but it was not the determining criteria.

The Knight lab has successfully used GitHub for several years for both code hosting and automated testing, without noticing any major deficiencies for its CPU-only software projects. However, with the recent porting of some codes to GPUs [4], we were unable to execute test suites automatically on pull requests with the freely available test environment.

Fortunately, GitHub offers a solution in the form of self-hosted **Action Runners** [5]. GitHub provides the necessary executables and authentication credentials that project maintainers can deploy on hardware under their control, and GitHub Actions will transparently schedule the automated tests on that external system as needed.

We had access to the Kubernetes-based Prototype National Research Platform (PNRP) Nautilus [6] system, so we investigated ways to dynamically deploy the Action Runners as Kubernetes pods [7]. Deploying such a setup on a dedicated node, while trivial, is wasteful as our automated tests rarely get executed more than once a week. Unfortunately, although we found several readily available Kubernetes integrations [8], they all rely on having cluster-admin level privileges. Obtaining cluster-admin level Kubernetes privileges in multi-user environments is a high bar for us and other users of Github, so we sought a solution that did not require elevated permissions. We thus decided to develop our own solution, which resulted in a lightweight and relatively generic solution.

## 2 THE UNPRIVILEGED KUBERNETES RUNNER MANAGER

As mentioned in the previous section, deploying a runner is trivial, both on bare metal and in a container. The crucial part we were missing was dynamic provisioning capability.

The approach we took is based on a polling model and implemented as a persistent Python-based service [9], instantiated as a Kubernetes deployment-managed pod. The process periodically queries GitHub Actions for outstanding jobs [10] and dynamically adjusts the number of runner pods accordingly. We limit ourselves to at most one runner at any time for simplicity.

The runner pod itself is also deployment-managed, with a fully defined configuration at deployment creation time. This allows the management pod to change the replication count [11], without worrying about other details. However, due to GitHub Actions design, modest sized persistent storage must be available to the runner pod and one-time initialization is needed.

Since the management pod is lightweight, requiring only a small amount of memory and short spikes of single core CPU compute, and the deployment objects being essentially free, the ongoing cost of this solution is minimal and can easily blend in the background compute of any non-trivial Kubernetes cluster. The runner pods can of course be as demanding as needed for the test jobs, but since they are active only while the test jobs are running, there is no waste associated with them.

### 2.1 GitHub Actions token lifetime considerations

The Action Runner security is based on a GitHub issued token, that is requested by a project maintainer, and is then used to establish an initial trust. The token is periodically, and automatically, renewed while the runner has a communication channel established with GitHub servers. With the expectation that the runner will be active at all times, to minimize security risks, token lifetimes are set to one week.

However, this can be a problem for an on-demand runner setup with projects that have infrequent pull requests, as there is no guarantee that a runner will be needed within the token lifetime. To ensure a valid token, we thus added logic to the manager to forcefully provision a runner at least twice a week, even if no eligible jobs are available, to ensure a valid token. While this will result in a small amount of underutilized resources, we believe it is an acceptable compromise.

### 2.2 The manager logic

The manager logic is a simple endless loop, with proper integration against both Kubernetes and GitHub. Figure 1 provides a schematic overview.

The manager needs two credentials. We authenticate with Kubernetes using a native, namespaced service account token [12] that is automatically propagated into the pod. We authenticate with GitHub using a fine-grained personal token [13], belonging to a project manager, and pass the GitHub token to the pod using a Kubernetes secret.

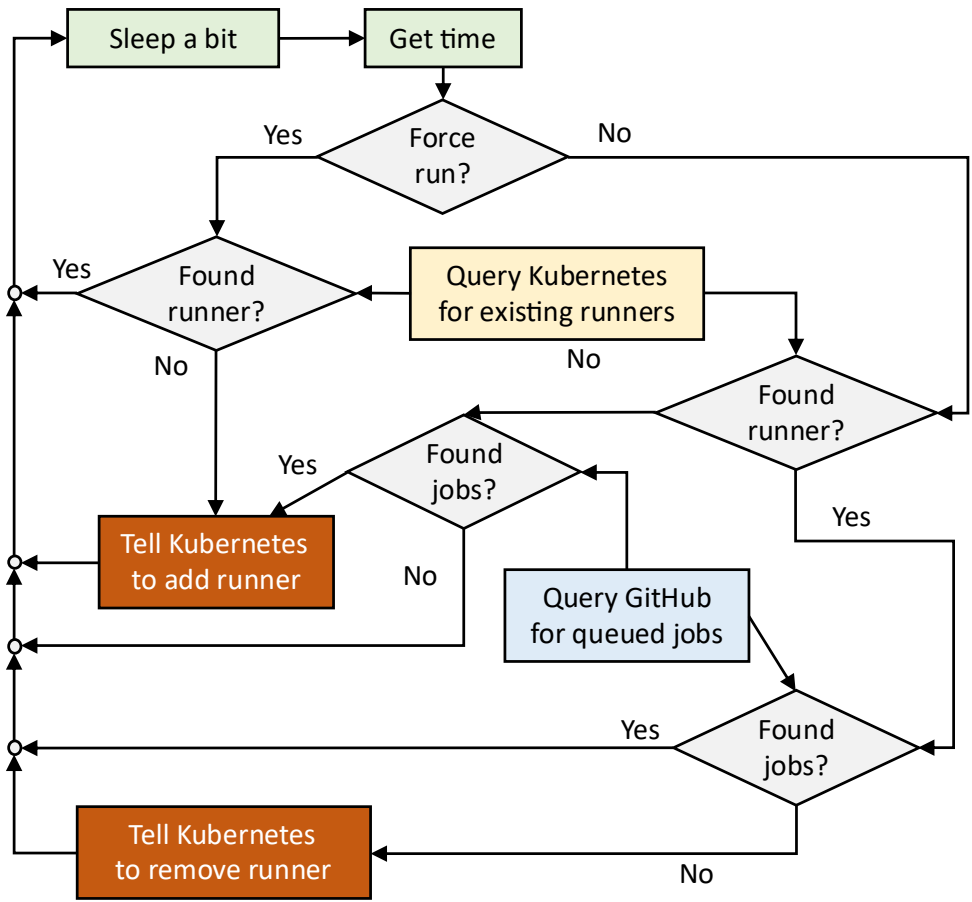

Figure 1: Schematic overview of the runner manager logic

### 2.3 The runner pod setup and operation

The manager pod does not put any hard requirements on how the Action Runner pod is set up. The only assumption made is that it is fully configured and will successfully connect and authenticate with GitHub.

That said, we had to settle on a specific configuration. In our case, the pod container image [9] is based on a CUDA-enabled version of Ubuntu as our primary goal was testing GPU-enabled code. This container image includes all the system packages that the GitHub software requires. Several helper scripts are also included.

The GitHub provided software itself, alongside the needed configuration and log files, lives on an external persistent volume and is initialized once, using one of the provided helper scripts, during the provisioning service setup. This design choice is driven by the fact that the Action Runners software auto-updates as needed, so preserving those changes is highly desirable. Nevertheless, to achieve maximum performance, the runner is instructed to keep all test job files on a local ephemeral partition.

Another helper script is used as the runner pod entry point and will invoke the GitHub provided software stack. Once invoked, the GitHub-provided runner processes take over and behave as if they were running on a dedicated node.

There is no automatic shutdown of the runner from inside the pod itself. The de-provisioning is fully driven by the management pod, by requesting the runner pod termination. Since that is supposed to happen only when no test jobs are available, the GitHub-provided software stack deals gracefully with such events.

## 3  KNIGHT LAB EXPERIENCE USING THE RUNNER MANAGER

The Knight lab has been operating two Action Runner managers on the PNRP Nautilus system for several months, one for the biocore/unifrac-binaries and one for the biocore/unifrac GitHub repositories. A single, dedicated Kubernetes namespace was created for our group, and the two management instances were deployed in that namespace. No special privileges were requested or granted.

Nautilus provides various persistent storage options [14]. We picked the POSIX-compliant shared CephFS solution, since it is available on all the nodes, allowing for maximum scheduling flexibility. It should be noted that Nautilus-provided CephFS is not a high-performance solution, but it is more than adequate for hosting the runner code and the associated config and log files. For convenience, we provisioned a single 20 GB shared partition for our group, with the two manager pods residing each in its own dedicated directory tree. At the time of writing, each manager pod used less than 1 GB on the shared filesystem, spread over about 18k unique files.

All nodes on Nautilus also provide NVMe-based local ephemeral storage, so the runner pod makes a request for 80 GB of ephemeral storage and makes it available for the test jobs. This amount is comparable to what is currently available on the native GitHub provided resources. We also provision a comparable number of compute and memory, i.e., 4 CPU cores and 8 GB of RAM, along the additional 1 NVIDIA GPU device.

To distinguish themselves from the GitHub provided runners, our on-prem runners advertise a custom OS string, namely "linux-gpu-cuda". This allows for easy targeting of these runners in the GitHub Action's workflow YAML.

Note that the two repositories contain different code and thus run significantly different build and test scripts. One is building OpenACC-based C++ code, while the other is just packaging Python code, some of which connects to GPU-enabled shared libraries imported from external packages. Nevertheless, the validation tests for both repositories exercise both CPU and GPU code paths that may have been directly or indirectly affected by a pull request. Having a GPU-enabled system on which to automatically run them has thus been extremely valuable, avoiding lengthy and error-prone manual tests by project managers.

The experience so far has been mostly positive, with most test jobs starting within minutes and performing as expected. The only major encountered problem, since resolved, was the GitHub Actions token lifetime limit described in the previous section. The original implementation did not include the periodic forced provisioning of the runners, resulting in runners that could not authenticate to GiHub until manually re-initialized.

## 4  SUMMARY AND CONCLUSIONS

GitHub is a great versioning control service, but its native automated build and test capabilities are limited to only a subset of the compute resource types one may want to support. In particular, at the time of writing there was no support for GPU compute. Fortunately, adding custom resources is possible by means of self-hosted Action Runners.

The main issue for projects that have infrequent pull requests is dynamic provisioning of such self-hosted runners. The Knight lab team, in partnership with the San Diego Supercomputer Center, has devised a lightweight solution that integrates with the Kubernetes-based PNRP Nautilus system and has been successfully using it for several months in support of two of their repositories.

The developed solution does not require any elevated privileges, unlike most other Kubernetes integrations. And while it has been optimized for and tested only on the PNRP Kubernetes environment, its design is general and should be usable on many other Kubernetes systems. We believe that this work opens new avenues for Github projects with specialized hardware or testing needs, both using GPUs and other specialized hardware types.

## ACKNOWLEDGMENTS

This work was partially supported by the National Institutes of Health (NIH) award U19AG063744 and National Science Foundation (NSF) grant OAC- 2112167.